# The Significance of the Cosmic Virial Theorem

James G. Bartlett and Alain Blanchard

Observatoire astronomique de Strasbourg, U.L.P., 11, rue de l'Université, F67 000 Strasbourg, FRANCE
Unité associée au CNRS

June 20, 1995

**Abstract.** We discuss the development and physics of the cosmic virial theorem and its traditional application as an indicator of the cosmic mean density. The standard result that the density must be sub-critical relies on the assumption that the galaxy three-point correlation function accurately describes the mass distribution around a typical pair of galaxies. To challenge this assumption, we develop a simple model of this mass distribution based on the extension of galactic halos. In such a model, one sees clearly the importance of the local mass distribution around pairs to the prediction of the galaxy pair-wise velocity dispersion. The model provides insight into the mechanics of the theorem and demonstrates that a flat universe is *consistent* with the observations. More generally, we conclude that the relative pair-wise velocity dispersion probes the mass clustered around galaxies and that the observations, viewed in this light, indicate a lower bound of $\sim 300\ h^{-1}$kpc on the radius of galactic halos.

## 1. Introduction

Since its inception (Peebles 1976a,b), the so-called 'cosmic virial theorem' (CVT) has been considered one of the most reliable indicators of the cosmic mean matter density, and for good reason. The theorem is a statistical statement of the virial theorem applied to the entire galaxy distribution and therefore does not suffer from biases associated with identifying individual, bound groups and clusters of galaxies. In this context, the two point correlation function $\xi(r)$ acts as a measure of the potential energy of the system, while the dispersion in the velocity difference of pairs of galaxies, $\sigma(r)$, represents the kinetic energy. Thus one understands the nature of the relation:

$$\sigma_{12}^2 \sim \Omega \xi(r) r^2, \tag{1}$$

where the density parameter $\Omega$ ($\equiv 8\pi G\bar{\rho}/3H_o^2$, where hereafter the Hubble constant $H_o = 100h$ km/s/Mpc) enters to normalize the potential energy associated with the clustering described by $\xi(r)$. Both $\sigma(r)$ and $\xi(r)$ are readily measured using a catalog of galaxy redshifts. The resulting determination of $\Omega$ is free of the problems inherent to cosmological distance indicators, which is the beauty of the approach.

The theorem (a term used loosely here) was conscripted immediately (Peebles 1976a,b), but not with convincing results until the completion of the first CfA redshift survey (Davis & Peebles 1983). Based on their observed galaxy correlation function and a deduced pair-wise velocity dispersion of $\sim 350$ km/s, Davis and Peebles concluded that $\Omega \sim 0.2$, a result more or less in agreement with applications of the CVT to other data sets (Bean et al. 1983; Hale-Sutton et al. 1989; Fisher et al. 1993b). Such a low value of the mean density affronts the theorist's preference for a flat Universe and, in particular, poses a problem for the 'standard' cold dark matter scenario (CDM) (Peebles 1982; Blumenthal et al. 1984).

The difficulty was clearly seen in early numerical simulations of the CDM model (Davis et al. 1985) where the particle pair-wise velocity dispersion was found to be $\sim 1200$ km/s when the particle correlation function was evolved to match the amplitude of the observed galaxy correlation function. The concept of biased galaxy formation (Bardeen 1986; Kaiser 1986) helps to alleviate this problem. In the biased picture, galaxies do not trace the underlying mass distribution, so that one may distinguish the mass correlation function from the galaxy correlation function. The traditional view was that galaxies formed only on high amplitude peaks of the density field and were therefore more strongly correlated than the average particle distribution. This can be represented by writing

$$\xi_{gg}(r) = b^2 \xi_{mass}(r)$$

at a certain scale $r$. Bardeen et al. (1986) found that in fact this relation holds reasonable well, with $b$ independent of $r$, for gaussian perturbations in the linear regime. Given the results of Davis & Peebles (1983) and remembering our simple relation 1, it would seem that a $b \sim 2$ is needed to allow an $\Omega \sim 1$. It appeared that the same amount of biasing reproduced the shape and amplitude of the galaxy correlation function in numerical simulations and could account for the observed abundances of galaxies and galaxy clusters (Davis et al. 1985; Bardeen et al. 1986).

With increased computer power and consequent numerical resolution, it became possible to study galaxies identified as dark matter *halos* in the simulations rather than as single mass points. Some interesting results followed. The clustering of the halos was found to be biased relative to the general mass distribution, supporting the biasing anzatz (White et al. 1987). This result can also be seen using analytical arguments (Schaeffer 1987; Bernardeau & Schaeffer 1992). More recently, Carlberg in their *spatial* clustering, dark matter halos exhibit a *velocity* bias in the sense that their velocities are typically smaller than that of the average mass particle. Unfortunately, the origin of this effect is not well understood and its magnitude depends upon the halo identification algorithm, although its existence in the simulations would seem to be well established (Cen & Ostriker 1992).

Until the detection of temperature fluctuations in the Cosmic Microwave Background (CMB) by the COBE satellite (Smoot et al. 1992; Wright et al. 1992), the biasing scheme seemed to many to be an acceptable solution to the problems specific to the CDM model and, more generally, to any flat cosmogony. However, the amplitude of the CMB fluctuations implies that mass fluctuations on the scale of 8 $h^{-1}$ Mpc must be the same as those observed in the distribution of optically selected galaxies; in other words, the model must be unbiased: $b_{optical} = 1$. Somewhat earlier, Peebles et al. (1989) pointed out that a single value for the bias parameter does not necessarily lead to a consistent picture, that the effects of biasing are not so simple to interpret, and that perhaps the pair-wise velocity dispersion may only be reduced by $1/b$ relative to an unbiased model. As we shall shortly see, the issue is subtle for the pair-wise velocity dispersion because it is not just the two-point function entering into the CVT, but also the three-point correlation function. Thus, the question remains, what to make of the low density indicated by the CVT?

In this paper we will to discuss this issue in a phenomenological fashion without recourse to the framework of a full theory of galaxy formation; an approach we feel is particularly appropriate given the present state of confusion over which theory(ies) are favored by which data. We begin in the following section with an examination of the physics underlying the CVT and identify some reasons for which we might suspect the deduced value of $\Omega$. We then briefly outline the full development of the CVT and discuss the traditional application of the theorem. This involves adopting the galaxy correlation functions to describe the actual mass distribution around galaxies. It is this assumption that we challenge in this *paper*. In section III we construct a simple, but we believe appealing, model of the mass distribution around galaxies. The model is cosmologically flat *and* consistent with the CVT, demonstrating the important weakness of the CVT: our lack of knowledge of the mass distribution. In this sense, the CVT may be giving us more information about this mass distribution around galaxies than about the cosmological mean mass density. This consideration will also lead us to conclude that galaxy halos extend out to at least $300 h^{-1}$ kpc.

## 2. Development and Application of the CVT

*2.1. Physical significance of the pair-wise velocity dispersion*

In order to discuss the implications of the galaxy pair-wise velocity dispersion, we must careful study the mass distribution around galaxies. Unfortunately, we do not have any direct probe of this distribution, other than the velocity dispersion itself. On the other hand, the galaxy distribution is well known to be amply described by a power-law two-point correlation function from 10 kpc to about 10 Mpc:

$$\xi(r) = \left(\frac{r}{r_0}\right)^{-\gamma}, \quad \text{with } r_0 \approx 5.5 h^{-1} \text{Mpc and } \gamma \approx 1.77, \quad (2)$$

where the parameters given correspond to optical galaxies (Davis & Peebles 1983). In addition, it seems that the general N-point galaxy correlation function, $\xi_N$, is also a simple power law and can be related to the two-point function by (White 1979; Alimi et al. 1990; Szapudi et al. 1992)

$$\xi_N = \xi^{N-1}.$$

This power-law behavior appears in other quantities, such as the cluster-cluster and cluster-galaxy correlation functions, all of which suggests the absence of any preferred scale in the galaxy distribution (although it does not necessarily imply that the galaxy distribution can be considered a fractal [Peebles 1993]).

For our present purpose, we are looking for a description of the *galaxy-mass* correlation function, and, from the above considerations, it would seem reasonable to assume that it too can be modeled as a simple power-law. On very short scales, we possess the rather direct probe of the galaxy-mass correlation function provided by galaxy rotation curves. Radio observations indicate that the rotational velocity $V_c$ of a typical spiral galaxy remains constant out to a radius on the order of 100 kpc; thus, on these scales, we may indeed write the density profile as a power-law:

$$\rho_g(r) = \frac{V_c^2}{4\pi G r^2} \propto 1/r^2.$$

Note that the exponent is interestingly close to that of the galaxy correlation function. Now, suppose that light traces mass, i.e.

$$\rho(r) = \overline{\rho} \xi_{gg}.$$

If we further assume that $\rho(r)$ is continuous, we would equate, at a certain radius $R$, this expression with the previous expression for the typical halo profile:

$$\overline{\rho} \xi_{gg} = \Omega \frac{3 H_0^2}{8\pi G} \left(\frac{R}{r_0}\right)^{-\gamma} = \frac{V_c^2}{4\pi G R^2}.$$

An average $L_*$ galaxy has a circular velocity of about 220 km/s at $R = 10$ kpc, from which we find an $\Omega \sim 0.5$. This large a mean density is somewhat surprising given the assumption that galaxies trace the mass.

Next, consider the pair-wise velocities of galaxies in this simple scheme. If one member of the pair is a $L_*$ galaxy, the second member is likely to be less luminous and may therefore be regarded as a satellite of the first. In this case it is legitimate to estimate the pair's relative velocity as the velocity caused by the mass distribution around the larger galaxy:

$$\sigma_{12}^2(r) = \frac{1}{2} \frac{GM}{r},$$

where $\sigma_{12}$ is the *line-of-sight* velocity difference of a motion deemed to be isotropic. Approximating, as we have, the mass distribution from 10 kpc and outwards by a power-law with index $\gamma$, we find that the pair-wise velocity increases slightly with distance:

$$\sigma_{12}^2(r) \approx \frac{1}{\gamma} (220 \, \text{km/s})^2 \left(\frac{r}{10 \, \text{kpc}}\right)^{2-\gamma},$$

This is quite consistent with the observed values.

One may argue that, since we have assumed that the second galaxy was much lighter than the first, this calculation underestimates the pair-wise velocity dispersion. However, if one assumes that the two galaxies have the same mass, the relative pair-wise velocity increases only by a factor $\sqrt{2}$ and remains consistent with the observed value. Furthermore, it is not clear that this correction should be made: Note that in real, magnitude-limited catalogs, galaxy luminosities encompass a wide range values, and therefore pairs with two $L_*$ galaxies are rare and do not dominate the statistics.

Another limitation to this simple calculation is that the influence of neighboring, third-party galaxies is not considered. We may easily account for this. At a distance $R$, the mean number of such correlated neighbors is roughly given by

$$N_c = \int_0^R \xi(r) n_g dV \approx \left(\frac{R}{0.65 h^{-1}\,\mathrm{Mpc}}\right)^{3-\gamma}.$$

When several galaxies are involved, we may assume that they all experience the effects of a single halo created by their individual contributions. The pair-wise velocity dispersion is then expected to be

$$\sigma_{12}^2(R) \approx (1 + N_c)\frac{GM}{R} \propto R^{5-2\gamma},$$

reaching 410 km/s at $1 h^{-1}$Mpc. This is slightly larger than the observational determinations, but not by much.

This simple view of the mass distribution provides a useful and heuristic description of the origin of the pair-wise velocity dispersion and of the physics underlying the CVT. It also highlights an important aspect of the interpretation of the pair-wise velocity dispersion: An accurate description of the mass profile around galaxies out to scales on the order of a Mpc is needed. This is demonstrated by the reliance of our immediate results on the assumption of continuity and by their dependence on the radius $R$. The usual approach to the CVT ignores any possible relation between the halo profiles and the large scale mass distribution, and yet data on the pair-wise velocity dispersion connects scales as small as 10 kpc, clearly galactic scales, to those are large as a few Mpc.

## 2.2. The standard cosmic virial theorem

As expected given its name, the CVT expresses the balance between kinetic and potential energies for sufficiently relaxed scales of the clustering hierarchy. The fundamental formula descends from the second equation of the BBGKY hierarchy (Peebles 1980), an approximation scheme for the Louiville equation, as applied to the cosmological mass distribution:

$$\frac{\partial [\xi(r)\sigma^2(r)]}{\partial r} = -\frac{2G\bar{\rho}}{r} \int d^3 z \zeta(r, z, |r-z|) \frac{r \cdot z}{z^3}. \quad (3)$$

In this equation, $\xi(r)$ is the particle two-point correlation function at separation $r$, $\sigma(r)$ is the pairwise velocity dispersion of particles along their common axis, $\bar{\rho}$ is the mean cosmological mass density, and $\zeta$ is the particle three-point correlation function. The right-hand side represents the gravitational acceleration of a pair of particles, separated by distance $r$, towards each other due to the surrounding mass. This is equated to the derivative of the velocity dispersion, which in the present context serves as the pressure term needed to maintain equilibrium against this acceleration. Thus, the theorem is best understood as an equation of hydrostatic equilibrium.

To reduce the the second BBGKY equation, an exact relation, to the given form, one must assume that the clustering is strongly nonlinear, so that the dominant gravitational term is due to the three-point function, and that it is stable, allowing one to drop time dependent terms. In essence, this latter point is the assumption that the scale $r$ has relaxed out from the Hubble expansion and moves through stages of quasi-static equilibrium as the universe expands. This condition would be manifest by a *mean* pair infall velocity exactly opposed to the Hubble velocity at separation $r$. In numerical simulations, this does not appear to be the case at $rh = 1$ Mpc, the radius at which the CVT is usually applied. Instead, mean infall velocities are about 4 times the Hubble rate (Fisher et al. 1993b; Zurek et al. 1993). However, this suggested lack of perfect stabilization does not invalidate the CVT: As long as the rate of the general collapse is slower than the rate at which equilibrium can be established, the CVT should hold. The simulations show a general infall velocity which is typically several times smaller than the velocity dispersion. Perhaps, then, stable clustering is not a bad approximation.

Adopting equation 3 as it stands, one is left with the question of how to evaluate the two terms and thereby determine the mean cosmological density. In redshift-space, the two-point correlation function is distorted along the line-of-sight by the relative motions of the galaxies (Peebles 1980). Fitting the distortion yields both the true spatial two-point function and the pair-wise velocity dispersion (see Fisher et al. 1993a for an excellent discussion of the procedure). The two-point function of both optical and IRAS selected galaxies is well described by a power law, equation 2, over scales from $rh \sim 10$ kpc to $rh \sim (10 - 20)$ Mpc, although for the IRAS galaxies one finds $hr_o = 3.76$ Mpc and $\gamma = 1.28$ (Fisher et al. 1993b). Additionally, both angular and redshift galaxy catalogues have demonstrated that the galaxy three-point function may be written as a cyclic permutation of terms involving the square of the two-point function:

$$\zeta(r, z, |r-z|) = Q[\xi(r)\xi(z) + \xi(r)\xi(|r-z|) + \xi(z)\xi(|r-z|)], (4)$$

$Q$ being the proportionality constant (Groth & Peebles 1977). Using counts in cells, Gaztañaga (1992) finds $Q = 0.7$ for the CfA catalogue, while Bouchet et al. (1993) find $Q = 0.5$ for the IRAS 1.2 Jy survey. It should be noted that this latter value is reportedly sensitive to the weight, or lack thereof, given to clusters in the IRAS sample.

Putting equations 2 and 4 into the integral of the CVT, we see that the first term vanishes by symmetry and that the second is easily evaluated using Gauss' theorem. The third term produces an integral with a large contribution near $z = 0$, which means that the galaxy at $z = 0$ receives an appreciable acceleration from the mass immediately surrounding it. Peebles (1976a,b; 1980) has removed this unwanted effect by either softening the gravitational force at small separations or by introducing a core radius into the otherwise pure power-law two-point function to account for the finite size of galaxies. Using the latter approach and taking the core size to be much smaller than the pair separation yields the following, simplified expression (Peebles 1980):

$$\sigma^2(r) = \frac{3}{4}\Omega \frac{QJ(\gamma)}{(\gamma-1)(2-\gamma)(4-\gamma)}(H_o r)^2 \xi(r),$$

where $J(\gamma)$ is a dimensionless integral (Peebles 1980, eq. 75.15 and 14). Note that this equation has the form proposed in the introduction.

From a redshift catalog, one may determine all the various components of this expression and therefore find $\Omega$. For example, if we use the correlation function of the CfA data ($\gamma = 1.77$ and $r_o = 5.5 h^{-1}$ Mpc), we find

$$\sigma^2(r) = \Omega(1000\,\text{km/s})^2 \left(\frac{Q}{0.7}\right)\left(\frac{rh}{\text{Mpc}}\right)^{0.2}. \qquad (5)$$

Implicit in this calculation is the assumption that the relative motion of galaxy pairs in the relaxed regime is isotropic. This permits us to equate $\sigma(r)$ in the CVT, representing the pair-wise velocity dispersion along the common axis, to the observable *line-of-sight* velocity dispersion. At $rh = 1$ Mpc, Davis & Peebles (1983) found $\sigma(r) \sim 350$ km/s, which demands that $\Omega \sim 0.1$. This is the well known result that the CVT requires an open universe. The velocity dispersion predicted by equation 5 is shown in figure 1 for both an open and a flat universe. We also indicate the recent IRAS velocity determination at $rh = 1$ Mpc, which Fisher et al. (1993a), together with their IRAS correlation function, used to find a somewhat larger $\Omega = 0.38 \pm 0.24$. It is important to note that not all determinations of the galaxy pair-wise velocity dispersion come in with such low values; in particular, the recent analysis of the CfA and SSRS surveys indicates a value for $\sigma(rh = 1\,\text{Mpc}) \sim 540$ km/s (Marzke et al. 1995). In addition, Mo et al. (1993) and Zurek et al. (1994) have emphasised the importance of sample variance in the existing catalogs.

Note that although the correlation function of the IRAS galaxies has a lower amplitude than that of the optical galaxies, the velocity dispersion in the two catalogues is the same. This is the cause of the larger estimated $\Omega$ extracted from the IRAS data. It also highlights a worrisome point, namely, that we do not know what the true mass distribution is. If we assume that galaxies trace the mass, then which correlation function should we use? Many would argue that the IRAS correlation function is the better choice. But, is this valid?, and is it an accurate enough description of the mass for our purpose?

## 3. Interpretation

In this section we pursue the answers to the questions posed above. A fundamental observation to this end is that the characteristic velocities around a galaxy, as measured on galactic scales by rotation curves (Freeman 1970; Rubin et al. 1985), on larger scales by galactic satellites (White et al. 1983; Persic et al. 1993; Zariitsky et al. 1993), and finally on even larger scales by the velocity dispersion of galaxy pairs, are all similar. The mass description used above gives us no enlightenment on this smooth continuation of the galactic velocity profile. We should expect to see some feature in this curve signifying the transition from scales dominated by individual galaxy halos to those dominated by the large-scale mass distribution. The choice of a small $\Omega$ to explain the data requires a conspiracy between the the large-scale mass distribution and the average halo distribution to avoid displaying this *à priori* expected feature. In addition, as we have seen, the mass in the immediate proximity of a galaxy produces a large contribution to the pairwise velocity dispersion. We therefore desire a precise modeling of its distribution.

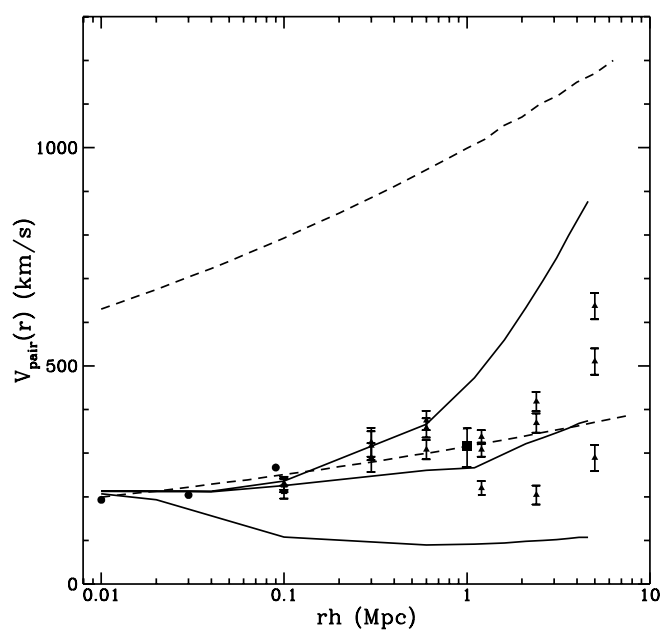

**Fig. 1.** The pair-wise velocity dispersion as a function of physical separation. The sets of triangles are the data from Davis & Peebles (1983) for values of $F = 0.1, 1$ and $1.5$ (see their paper for a discussion of this parameter). The three filled circles are the Davis & Peebles interpretation of the Turner (1976) galaxy pair catalog; the square represents the IRAS value at $1 h^{-1}$ Mpc (Fisher et al. 1993b). Equation 5 produces the upper dashed curve for a flat universe and the lower dashed curve for $\Omega = 0.1$. Adopting the model discussed in the text, a flat universe will produce the uppermost solid curve. The remaining two curves show the results of this same model for $\Omega = 0.1$ (middle solid curve) and $\Omega = 0.01$ (lowest solid curve). All of these results were obtained with $Q = 0.7$.

To illustrate the importance of these issues, we now develop a simple model of the cosmological mass distribution. Imagine that the galaxies cluster according to the observed correlation functions and that each is surrounded by dark matter with the profile of a singular, isothermal sphere out to some cutoff radius $r_c$. The quantity of interest in equation 3 is the *galaxy-galaxy-mass* three-point correlation function. On scales $r > r_c$ this is, as assumed previously, identical to the galaxy-galaxy-galaxy three-point function. However, on smaller scales the two are certainly *not* equal. In general, we can write the relation between these two correlation functions as follows:

$$\bar{n}^2 \bar{\rho} \zeta_{ggm}(r, z, |r - z|) = \bar{n}^2 \xi_{gg}(r)(\rho_g(z) + \rho_g(|r - z|))$$
$$+ \bar{n}^3 \int d^3 y\, \zeta_{ggg}(r, y, |r - y|) \rho_g(|z - y|), \qquad (6)$$

where the two galaxies are separated by $r$ and the mass point is at position $z$. The first two terms represent the contributions of each of the galaxy 'halos' to the mass at point $z$, while the integral adds the effect of a possible third galaxy at position $y$. We use the term 'halo' to mean the mass distribution around an individual galaxy. This should not be interpreted as a halo physically 'attached' to the galaxy, but rather as the *statistical* or *average* mass distribution around galaxies, as is appropriate when employing correlation functions.

dius: For separations $r < r_c$, the individual galaxy 'halos' play the dominant role while, on larger scales, these first two terms vanish and the integral dominates. This term can be reduced to the galaxy correlation function times the mean mass surrounding an individual galaxy. Then the galaxy-galaxy-mass and the galaxy-galaxy-galaxy correlation functions in equation 6 become one and the same. The flatness of the observed velocity profile, $\sigma^2(r)$, can be understood in one of two ways in this model: either as a conspiracy of the integral term and the two 'halo terms' to maintain the flat velocity profile or as an indication that the 'halo' size is in fact $> 1h^{-1}$ Mpc in extent.

mass three-point function as calculated with equation 6. All the features displayed by this curve are easily understood. On small scales the characteristic 'halo' velocity dominates the pair-wise velocity dispersion of galaxies, while on scales $r > r_c$ we begin to recover the previous value for $\Omega = 1$. The expected feature at the 'halo' cutoff appears as the smooth continuation between these two regimes.

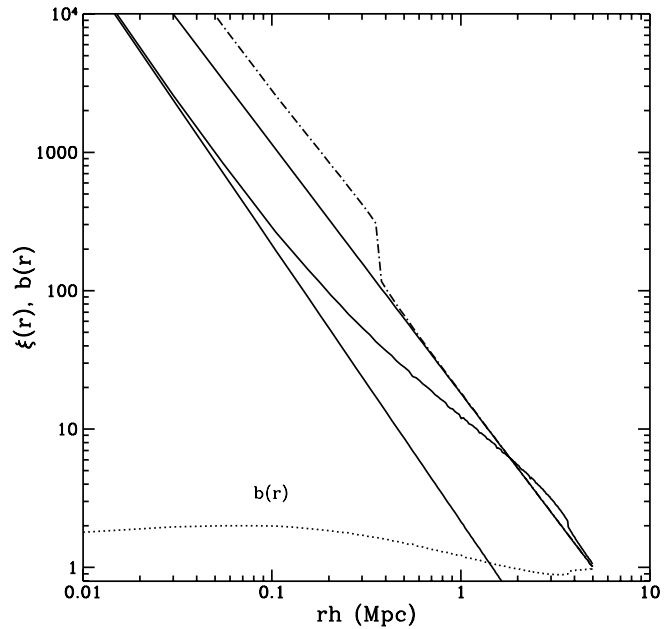

**Fig. 3.** The two-point correlation functions and the bias parameter $b(r)$. The two solid, straight lines represent the halo mass two-point function (the farthest to the left) and the galaxy-galaxy two-point function from the CfA catalog (straight line to the right). The solid curve shows the *galaxy-mass* correlation function of our model for $\Omega = 1$ smoothly joining these two. The effective bias in this flat model is given as a function of separation by the dotted line. Finally, the dot-dashed line is the galaxy-mass two-point function in our model for $\Omega = 0.1$.

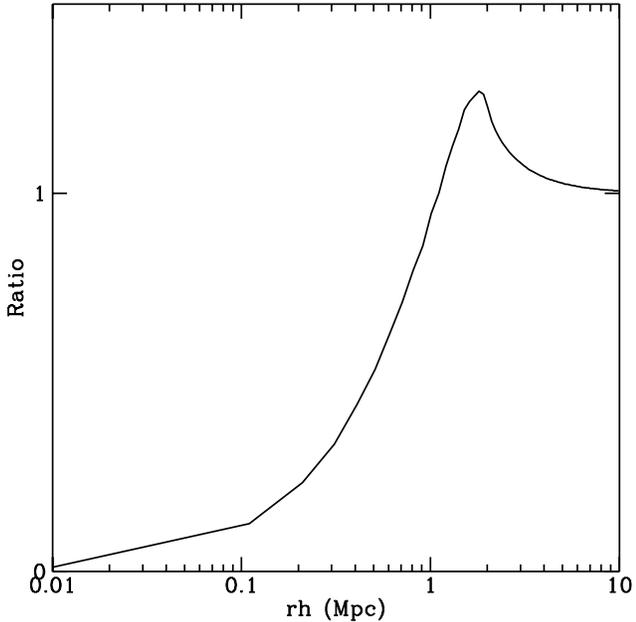

**Fig. 2.** The ratio of the integral term in equation 6 to the galaxy three-point correlation function. The parameters chosen here are those of the CfA survey (Davis & Peebles 1983) ($\gamma = 1.77$ and $r_o = 5.5h^{-1}$ Mpc). The ratio is independent of $Q$.

To expand on this second viewpoint, consider a universe in which all the mass clusters in the 'halos' surrounding galaxies. The halo size or cutoff radius $r_c$ is related to the galaxy number density $n_g$ and characteristic circular velocity $v_g$ by

$$r_c h = (2.98\,\mathrm{Mpc})\Omega \left(\frac{0.01 h^3\,\mathrm{Mpc}^{-3}}{n_g}\right)\left(\frac{200\,\mathrm{km/s}}{v_g}\right)^2.$$

In a flat cosmology we indeed have the second situation of a large $r_c$, implying that the mass distribution is different from the galaxy distribution at $\sim 1$ Mpc. In figure 2 we show the ratio of the integral term in equation 6 to the galaxy three-point correlation function. On large scales the ratio approaches unity as expected, but, moving inwards, the extended 'halos' reduce the mass correlations relative to those of the galaxies. Therefore, using the galaxy correlations to model the mass correlations results in a serious overestimate of the predicted pair-wise velocity dispersion. In figure 1 we show the CVT-predicted pair-wise velocity dispersion in this scenario using the correct

The remaining two solid curves in figure 1 show the predicted velocities in our toy model for lower values of $\Omega$. We see no evidence of a break in the velocity curve for the $\Omega = 0.1$ case, the lack of which represents a conspiracy between the various terms in equation 6. For an even lower cosmic density, an actual decrease in the predicted velocities appears beyond the 'halo' extent. The trio of curves nicely demonstrates the physics of the CVT.

To better understand the model, we calculate the *galaxy–mass* two–point function $\xi_{gm}$:

$$\bar{n}\bar{\rho}\xi_{gm}(r) = \bar{n}\rho_g(r) + \bar{n}^2 \int d^3y\, \xi_{gg}\rho_g(|r-y|). \tag{7}$$

Although not of immediate use, for it is rather the galaxy–galaxy two–point function entering equation 3, the result of this calculation illustrates well the nature of the mass distribution in our model. The solid curves in figure 3 correspond to a flat Universe. Here we see the smooth transition between the small scales, dominated by the halo, and the larger scales,

galaxy–mass two–point function approaches the galaxy–galaxy two–point function. The figure also shows the corresponding bias factor defined as $\sqrt{\xi_{gm}/\xi_{gg}}$. Note that the bias is not constant with scale and that the galaxies are anti–biased on intermediate scales (see also figure 2). This anti–bias is the result of mass conservation: In treating the galaxies as extended rather than point-like objects, we have exiled some mass from the smaller to the larger scales - the galaxies have been 'softened'.

For comparison, we present, as the remaining curve in figure 3, the galaxy-mass correlation function for an $\Omega = 0.1$ universe. In this case, the galaxies are anti-biased on small scales. Had the galaxies been point masses, the correlation function would be identical to the galaxy-galaxy function; however, as the universal mean density decreases in the model, the finite halo density demands that the mass correlation rise in order to account for these halo properties with a decreasing amount of mass. One easily sees this in the first term of equation 7. This once again illustrates the necessary appearance of the transition between the mass distribution internal to galaxies and the universal mass distribution on larger scales. It also further emphasizes the conspiracy required of an $\Omega = 0.1$ universe to explain the relatively flat observed pair–wise velocity curve: There is no feature in the velocity curve, but there is an abrupt transition in the galaxy-mass correlation function. It seems to us that these remarks apply independently of our simple model.

One may wonder whether the bias we find for a flat universe could explain in a simple way the model's low predicted pair–wise velocities. This is in fact not the case. A bias permits an upward correction to the $\Omega$ deduced from the standard CVT, equation 5, by a factor $b$ (Peebles et al. 1989). In applying this reasoning to our model of a flat universe, one would have to conclude that the bias was as large as 10, while the bias exhibited by the correlation functions is at most a factor of 2. A closer investigation shows that the result cannot be interpreted at all in terms of such a biasing scheme. The amplitude of the galaxy-galaxy-mass correlation function is not much different than the galaxy three-point function on a scale of one megaparsec (see figure 2); in fact, it is slightly larger. The reason for which the predicted pair–wise velocities are reduced is that the quasi-divergent term in the integral in equation 3, caused by the last term in equation 4, is replaced by the last term in equation 6. As we have seen in figure 2, this new term is much 'softer'.

## 4. Conclusion

We have explored the interpretation of the pair-wise velocity dispersion of galaxies by focusing on the main weakness of the cosmic virial theorem: the assumption that the galaxies are point-like and therefore trace the mass distribution. By adopting the observed galaxy correlations and assuming that each galaxy is accompanied by an extended 'halo', we developed a simple relationship between the galaxy and mass distributions. This is given in equation 6. We emphasize that this relation is statistical in nature, which perhaps alleviates some of the uncomfortableness of such a simple picture. In this scenario, the CVT (equation 3) predicts velocities in a flat universe as shown by the upper-most solid curve in figure 1. The data, out the $rh \sim 1$ Mpc, are well fit by the this curve. The model velocities rise rapidly beyond this point, but the validity of stable clustering may be questioned in this regime. Thus, we have a simple model of a flat universe with a pair-wise velocity dispersion in accord with the data. The origin of this result seems rather clear form our analysis: on scales up to $r_c$, the velocity dispersion only probes the extended mass distribution around galaxies, which also explains the flatness of the observed relative pair-wise velocity curve. This suggests that the galaxy pair-wise velocities should not be considered as *direct* evidence for a low density universe. On the contrary, from figure 1 we deduce, rather, a *lower limit* on $\Omega$ of $\sim 0.1$, since models with much smaller values predict a falling velocity curve. Given our model's relation between $r_c$ and $\Omega$, this implies that galaxy halos (or, as per our discussion, the mass distribution around galaxies) extend out to at least $\sim 300$ kpc. It is interesting to note that such an extension is also indicated by recent studies of quasar absorption-line systems (Sargent 1994; Lanzetta et al. 1995).